%% file: paper_revised1.tex
\documentclass[%
 reprint,
superscriptaddress,
 amsmath,amssymb,
 aps,
longbibliography
]{revtex4-2}

\usepackage{graphicx}
\usepackage{dcolumn}
\usepackage{bm}
\usepackage{xcolor}
\usepackage{mathrsfs}
\usepackage{soul}

\begin{document}

\title{Generating two-mechanical mode entangled cat states, and steady-state entanglement, in cavity optomechanics in the presence of dissipation}
\author{Sanket Das}
\email{sanket.das@oist.jp}
\affiliation{Quantum Machines Unit, Okinawa Institute of Science and Technology Graduate University, Okinawa 904-0495, Japan}
\author{Jason Twamley}
\email{jason.twamley@oist.jp}
\affiliation{Quantum Machines Unit, Okinawa Institute of Science and Technology Graduate University, Okinawa 904-0495, Japan}

\date{\today}
\thanks{A footnote to the article title}%

\begin{abstract}
We investigate a dissipation-engineering approach to produce a phase-dependent collective-mode Schrödinger cat state involving two modes. Our model features a single cavity mode that interacts with two spectrally identical mechanical oscillators. Both oscillators are coupled through a phase-dependent hopping interaction. We demonstrate that by adjusting the phase of the phonon-hopping interaction, one can control the bipartite entanglement of the phase-dependent two-mode cat state during its generation. Additionally, our study reveals that phonon interactions act as a tunable parameter, enabling the manipulation of steady-state entanglement between the bare mechanical modes, even in the presence of environmental effects and thermal excitations. Our scheme provides a feasible approach for the phase-dependent multi mode non-Gaussian states preparation. \\

\end{abstract}

\maketitle


\section{\label{sec:Introduction}Introduction}
Quantum states are fundamental elements for encoding, storing, and processing information in a quantum
system. Unlike classical information, this process uses a qubit, which is characterized by its basis states. A Schr\"{o}dinger cat state (SCS) \cite{Schrödinger1935}, a quantum superposition of two distinguishable macroscopic coherent states \cite{Schrödinger1926,Sudarshan1963EquivalenceBeams,Glauber1963CoherentField, Mehta1967DynamicsStates, Zhang1990CoherentApplications} serves as the logical basis for qubits. Recently, SCS has attracted attention for enabling bosonic quantum information processing, including quantum error correction \cite{Cochrane1999MacroscopicallyDamping,Li2017CatChannel,Gertler2021ProtectingCorrection}, bosonic encoding \cite{Cai2021BosonicCircuits}, fault-tolerant computation \cite{PhysRevX.9.041053}, and precision measurements \cite{PhysRevA.33.4033}. 

Motivated by these applications, considerable effort has been devoted to develop various techniques to generate SCS from a single bosonic mode. One such method involves a weak intensity-dependent frequency shift of a bosonic mode, known as Kerr nonlinearity-induced SCS generation \cite{Yurke1986GeneratingDispersion,Yurke1988TheDetection,Kirchmair2013ObservationEffect}. However, in most experimental platforms the Kerr nonlinearity is weak, leading to a long SCS generation time. Hence, the generated state becomes highly fragile due to the environmental decoherence. A parametrically driven Kerr oscillator helps overcome this, but it requires a strong Kerr interaction, which is experimentally demanding \cite{Grimm2020StabilizationQubit,He2023FastResonator} . Such a state has been investigated both theoretically and experimentally in optical \cite{Ourjoumtsev2006GeneratingProcessing,Ourjoumtsev2007GenerationStates,Sychev2017EnlargementStates,Glancy:08}, microwave cavities \cite{Mirrahimi2014DynamicallyComputation,Leghtas2013Hardware-EfficientProtection,Leghtas2015ConfiningLoss,Vlastakis2013DeterministicallyStates}, spin \cite{Agarwal1997AtomicStates,PhysRevA.87.052323}, and mechanical systems \cite{Bild2023SchrodingerOscillator}. Further investigation along the same direction suggests that parametrically exciting a bosonic mode with quadratic dissipation stabilizes the cat manifold at the steady state \cite{Mirrahimi2014DynamicallyComputation,Leghtas2013Hardware-EfficientProtection}. 

To realize two-phonon dissipation, it is essential to have multimode bosonic systems that feature nonlinear interactions and reservoir engineering. Cavity optomechanical systems (OMS) provide a platform for implementing such engineered dissipative dynamics \cite{Aspelmeyer2014CavityOptomechanics,
Kronwald2013ArbitrarilyDissipation,
Wang2013Reservoir-EngineeredSystems}. A nonlinear radiation-pressure interaction between a highly dissipative cavity and long-lived mechanical elements permits adiabatic elimination of the cavity \cite{Xu2015MechanicalSystems,Chakraborty2019DelayedPoints}, resulting in an engineered reservoir of mechanics \cite{Harrington2022EngineeredScience}. Reservoir engineering in cavity OMS has successfully demonstrated ground-state cooling of a mechanical motion \cite{Wilson-Rae2007TheoryBackaction,Genes2008Ground-stateSchemes,Teufel2011SidebandState,Chan2011LaserState}, mechanical squeezing \cite{Kronwald2013ArbitrarilyDissipation,Wollman2015QuantumResonator,Agarwal2016StrongDetection}, and entanglement between multiple mechanical modes in the resolved sideband regime \cite{Woolley2014Two-modeReservoir,Ockeloen-Korppi2018StabilizedOscillators,Tan2013Dissipation-drivenSystems}. Interestingly, these studies rely on a linearized radiation-pressure interaction, which is a beam-splitter interaction between the cavity and the mechanics. Consequently, the dynamical evolution only captures the Gaussianity of the states involved. One can lift the Gaussian nature of the states by considering a quadratically coupled OMS \cite{Asjad2014ReservoirDecoherence,Tan2013GenerationDissipation}. Despite nonclassical state preparation, it suffers from a weak single-photon coupling strength in a quadratically coupled OMS than the usual one. 

To overcome this limitation, researchers have theoretically predicted that the intrinsic nonlinearity of the radiation-pressure interaction is sufficient to generate SCS in a conventional OMS \cite{Hauer2023NonlinearMotion}. Incorporating one additional non-degenerate mechanical mode into the previous model predicts the generation of an entangled pair coherent state \cite{Yu2025DissipativelyStates}. In this work, we consider a couple of interacting degenerate, long-lived mechanical oscillators coupled to a common lossy cavity mode. A bichromatic drive is applied to the cavity mode. The adiabatic elimination of the cavity results in an effective nonlinear dissipative dynamics for the mechanical subsystem. We show that this engineered reservoir stabilizes two-mode mechanical SCS that depend on the phase of phonon hopping interaction. We further utilize the phase-dependent phonon-hopping interaction to tune the steady-state entanglement between the mechanical modes.

The paper is organized as follows. In Sec.  \ref{sec:theoretical model}, we describe our model and show how reservoir engineering modifies the dynamics of collective modes. Sec. \ref{sec:results} discusses two-mode cat state formation and bipartite entanglement between them in the presence of local damping. Sec. \ref{sec:bip_ent} analyzes the effect of intramechanical coupling on bipartite entanglement between the collective and bare mechanical modes at the steady state. Finally, we draw our conclusions in Sec. IV.

\section{\label{sec:theoretical model}Theoretical Model}
\begin{figure}[h]%
	\centering
	\def\svgwidth{0.8\linewidth}
    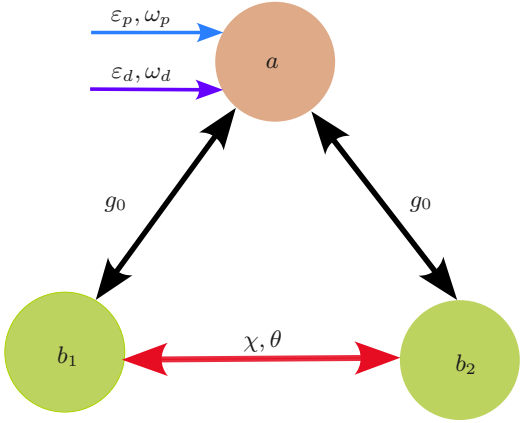
	\caption{\label{fig:fig1}{A schematic diagram of a cavity optomechanical system consists of a single cavity and two mechanical modes. The radiation pressure and phonon-hopping interactions are shown in black and red arrows, respectively. Single photon optomechanical coupling is $g_0$, and the phonon-phonon interaction is characterized by $\chi$ and $\theta$. Cavity is driven by the pump field of amplitude and phase $\varepsilon_p$ and $\omega_p$. Along with that, a coherent drive of amplitude $\varepsilon_d$ and frequency $\omega_d$ is also applied to the cavity.}}
\end{figure}

We consider a cavity optomechanical system that comprises of a single cavity mode interacting with two degenerate mechanical oscillator modes $b_1$ and $b_2$ through radiation pressure interaction, as shown in Fig. \ref{fig:fig1}. The mechanical modes in turn interact with themselves. We also drive the cavity with a bichromatic drive. The Hamiltonian of the composite system can be expressed as follows 
\begin{align}
\label{eq:eq1}
\frac{H}{\hbar}&=\omega_c a^\dagger a +\sum\limits_{i=1}^2\omega_{m_i}b_i^\dagger b_i +\sum\limits_{i=1}^{2} g_0 a^\dagger a(b_i^\dagger+b_i)\nonumber\\
&+\sum\limits_{i\not= j}\chi e^{i \theta_{ij}}b_i^\dagger b_j+\left(\varepsilon_p a^\dagger e^{-i\omega_p t}+\textrm{h.c.}\right)\nonumber\\
&+\left(\varepsilon_d a^\dagger e^{-i\omega_d t}+\textrm{h.c.}\right).
\end{align}
The first two terms of the Hamiltonian represent the free energy of each component of the system. The resonance frequencies of the cavity and the two movable mirrors are $\omega_c, \omega_{m_i}$ $(i=1,2)$, respectively. The third term represents the radiation pressure interaction between the cavity and each of the mechanical modes \cite{Das2022Phase-dependentResonator, Genes2008SimultaneousCavity, Huang2014DoubleMirrors}, characterized by single photon coupling strength $g_0$.  The fourth term depicts the phonon hopping interaction from $i$-th mechanical mode to $j$-th mechanical mode with coupling strength $\chi$ and phase $\theta_{ij}$. To consider synthetic gauge fields, we consider $\theta_{12}=-\theta_{21}=\theta$ \cite{Lai2022Noise-TolerantMagnetism, Lai2018SimultaneousOptomechanics,Lai2020NonreciprocalResonators,Xu2022MillionfoldCoupling}. The last two terms signify the interaction of the intracavity field with a bichromatic drive comprising a pump field with amplitude and frequency $\varepsilon_p~\textrm{and}~\omega_p$, respectively and a coherent drive characterized by amplitude and frequency $\varepsilon_d$ and $\omega_d$, respectively.
Now moving to a frame that rotates with pump frequency $\omega_p$, we obtain the fluctuating part of the rotated Hamiltonian as
\begin{align}
\label{eq:eq2}
&\frac{H_\textrm{f}}{\hbar}=-\Delta a^\dagger a + \sum\limits_{i=1}^2\omega_{m_i}b_i^\dagger b_i+\sum\limits_{i=1}^2g_1\left(a+a^\dagger\right)(b_i+b_i^\dagger)\nonumber\\
&+\sum\limits_{i\not= j}\chi e^{i \theta_{ij}}b_i^\dagger b_j+\sum\limits_{i=1}^2 g_0 a^\dagger a(b_i+b_i^\dagger)
+\left(\varepsilon_d a^\dagger e^{i\Delta_p t}+\textrm{h.c.}\right),
\end{align}
where $\Delta=\omega_p-\omega_c$ ($\Delta_p=\omega_p-\omega_d$) is cavity (coherent drive) detuning with respect to the pump frequency and $g_1=g_0\bar{\alpha}$ is the enhanced optomechanical coupling strength. Without loss of generality, we consider the steady-state intracavity field amplitude $\bar{\alpha}=|\varepsilon_p/(\Delta+i\gamma_a/2)|$ to be a real quantity. For large intracavity photon number, one can safely neglect the nonlinear interaction of Eq. \eqref{eq:eq2} \cite{Aspelmeyer2014CavityOptomechanics, Aspelmeyer2014CavityOptomechanicsb, Favero2014FocusOptomechanics}. However, in a bad cavity with a high damping rate, the intracavity photon number is low, making radiation-pressure-induced nonlinear interactions significant. At this stage, we incorporate the environmental effects on the cavity mode and each of the mechanical modes and the complete master equation reads
\begin{align}
\label{eq:eq3}
\dot{\rho}=&-\frac{i}{\hbar}[H_{\textrm{f}},\rho]+\gamma_a\mathcal{L}[a]\rho+\Gamma\left(n_b+1\right)\sum\limits_{i=1}^2\mathcal{L}[b_i]\rho\nonumber\\
&+\Gamma n_b\sum\limits_{i=1}^2\mathcal{L}[b_i^\dagger]\rho.
\end{align}
 The second and third terms represent the emission into their respective baths for optical and degenerate mechanical modes at temperature $T$ with mean thermal occupancy given by $n_b=\left(e^{\hbar\omega_m/k_B T}-1\right)^{-1}$, respectively. The fourth term accounts for absorption from mechanical reservoirs where $\mathcal{L}[O]\rho=O\rho O^\dagger-(1/2)\left(O^\dagger O\rho+\rho O^\dagger O\right)$ is the Lindblad superoperator of any operator $O$. When the cavity is a bad one, we adiabatically eliminate the fast-decaying cavity mode following a well-established technique of open quantum systems \cite{Gonzalez-Ballestero2024Tutorial:Systems,azouit2016adiabaticeliminationopenquantum,10.1143/PTP.20.948,Wilson-Rae2008Cavity-assistedResonators}. The reduced master equation for the mechanical modes is given in Appendix \ref{appendix1} 
\begin{widetext}
\begin{align}
\label{eq:eq4}
\dot{\rho}_s=&-\frac{i}{\hbar}[H_s,\rho_s]+\Gamma(n_b+1)\sum_{i=1}^2\mathcal{L}[b_i]\rho_s+\Gamma n_b\sum_{i=1}^2\mathcal{L}\left[b_i^\dagger\right]\rho_s
+\Gamma_i\mathcal{L}\left[\sum_{i=1}^2 b_i\right]\rho_s+ \Gamma_i'\mathcal{L}\left[\sum_{i=1}^2 b_i^\dagger\right]\rho_s+\Gamma_2\mathcal{L}\left[\left(\sum_{i=1}^2 b_i\right)^2\right]\rho_s \nonumber\\
&+ \Gamma_2'\mathcal{L}\left[\left(\sum_{i=1}^2 {b_i^\dagger}\right)^2\right]\rho_s,
\end{align}
\end{widetext}
where $\Gamma_i=4g_1^2\gamma_a/\left(\gamma_a^2+4(\Delta+\omega_m)^2\right)$ and $\Gamma_i'=4g_1^2\gamma_a/\left(\gamma_a^2+4(\Delta-\omega_m)^2\right)$ account for collective mechanical mode dissipation due to the lossy cavity. Furthermore, adiabatic elimination of the cavity induces a collective two-phonon decay rate $\Gamma_2=4|g_2|^2\gamma_a/\left({\gamma_a^2+4(\Delta+2\omega_m)^2}\right)$ and absorption rate $\Gamma_2'=4|g_2|^2\gamma_a/\left({\gamma_a^2+4(\Delta-2\omega_m)^2}\right)$ as delineated by the sixth and seventh terms, respectively. Along with modifying the environmental effects, it also transforms the coherent interaction as
\begin{widetext}
\begin{align}
\label{eq:eq5}
H_s/\hbar=&\sum\limits_{i\not= j}\chi e^{i \theta_{ij}}b_i^\dagger b_j+\varepsilon_2^*\left(\sum_{i=1}^2 b_i\right)^2+\varepsilon_2\left(\sum_{i=1}^2 b_i^\dagger\right)^2
+4|g_1|^2\left(\frac{\Delta-\omega_m}{\gamma_a^2+4(\Delta-\omega_m)^2}+\frac{\Delta+\omega_m}{\gamma_a^2+4(\Delta+\omega_m)^2}\right)\sum_{i,j=1}^2 b_i^\dagger b_j\nonumber\\
&+\frac{(\Delta+2\omega_m)|g_2|^2}{\gamma_a^2+4(\Delta+2\omega_m)^2}\sum_{i,j,k,l=1}^2 b_i^\dagger b_j^\dagger b_k b_l+\frac{(\Delta-2\omega_m)|g_2|^2}{\gamma_a^2+4(\Delta-2\omega_m)^2}\sum_{i,j,k,l=1}^2 b_i b_j b_k^\dagger b_l^\dagger.
\end{align}
\end{widetext}
The first term of $H_s$ represents the phase-dependent coherent interaction between two degenerate mechanical modes. The second and third terms explain a parametric drive acting on the collective mechanical mode with amplitude $\varepsilon_2= 2i\varepsilon_dg_2^*/\gamma_a$. Adiabatic elimination of the fast-decaying cavity generates a collective dispersive interaction as described by the fourth term. Additionally, the nonlinear radiation pressure gives rise to a self-Kerr interaction in the collective mode, as illustrated by the fifth and sixth terms of Eq. \eqref{eq:eq5}.\\
To understand the collective behavior of lossy cavity-mediated interactions, we move to the dark- and bright-mode bases. In the presence of interacting degenerate mechanical modes, the bright ($D_+$) and dark ($D_-$) modes are defined as 
\begin{align}
\label{eq:eq6}
D_{\pm}=&\frac{1}{\sqrt{2}}b_{1(2)}\pm\frac{1}{\sqrt{2}}e^{{\pm}i\theta} b_{2(1)},
\end{align}
respectively such that the commutation relations hold $\left[D_{\pm},D_{\pm}^\dagger\right]=1$ and $\left[D_+,D_-\right]=0$.
Utilizing the bright and dark modes, the effective master equation Eq. \eqref{eq:eq4} contains a phase-sensitive quadratic dissipator in the eigenmode basis. For a certain phase, it corresponds to a pair of excitation losses from either bright mode or dark mode. In contrast, other phase values result in a single excitation loss from each eigenmode in an indistinguishable manner, eventually leading to an excitation-pair loss. Consequently, the dissipation affects two eigenmodes simultaneously, producing a correlation between them. To quantify the correlation between the two eigenmodes, we evaluate logarithmic negativity to measure bipartite entanglement as
\begin{align}
\label{eq7}
\varepsilon_N = \textrm{log}_2||\rho^{T_A}||_1,
\end{align}
\begin{figure*}[!ht]%
	\centering
	{\includegraphics[width=\textwidth]{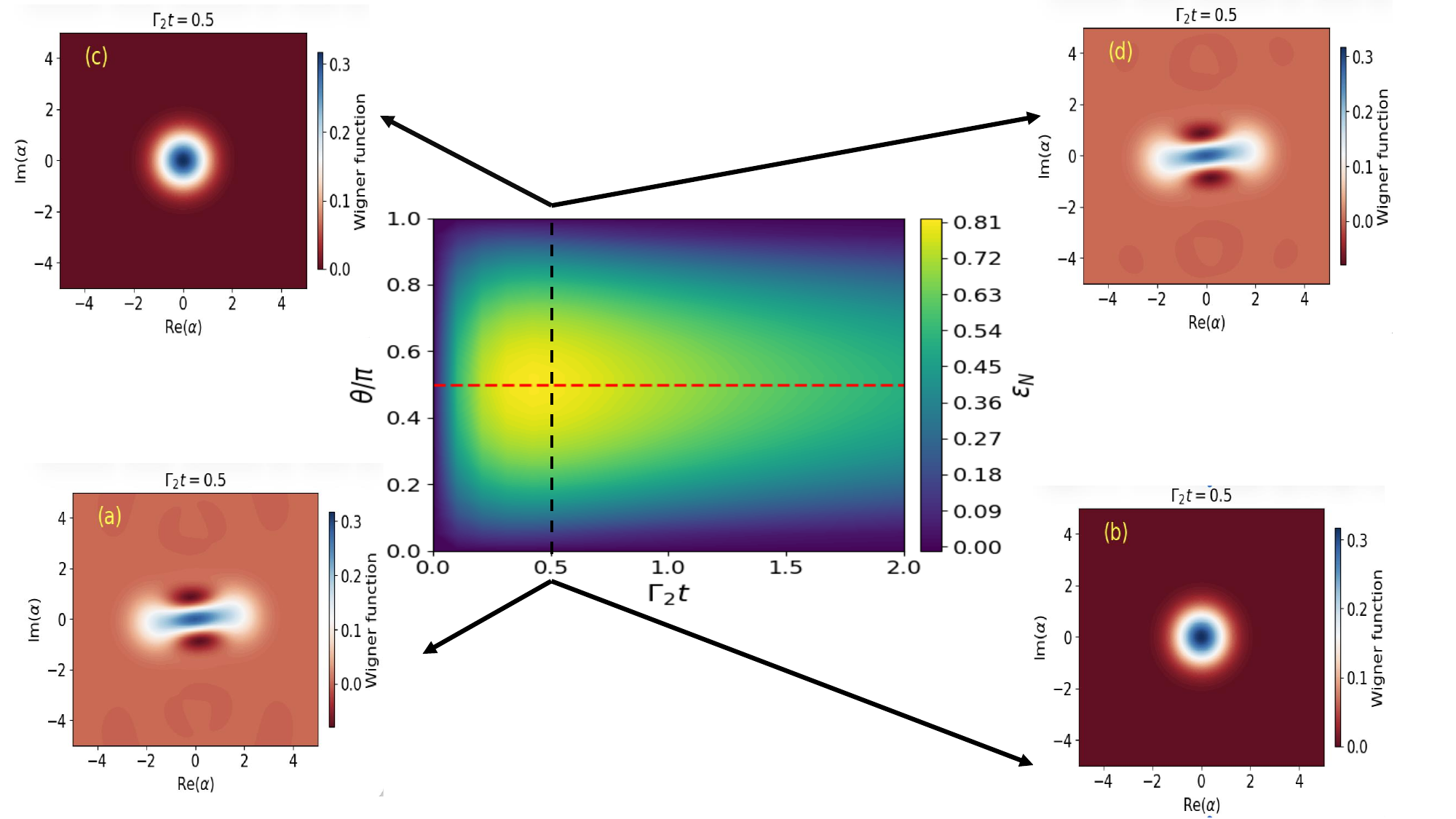}}
	\caption{\label{fig:fig2}{The central panel shows evolution of bipartite entanglement between the collective modes. The panels a(b) and c(d) depict the Wigner function of the bright(dark) mode when $\theta=0$ and $\pi/2$ at normalized time $\Gamma_2t = 0.5$, respectively. We consider the degenerate mechanical modes with resonance frequency $\omega_m/2\pi=15$ MHz and decay rate $\Gamma/2\pi=15$ Hz, respectively. Thermal occupancy of each mechanical mode $n_b=0$. The exchange interaction between the two mechanical modes is characterized by the interaction amplitude $\chi/\Gamma_2=10^{-2}$. The other parameters are $\Delta = -2\omega_m$, $\gamma_a/2\pi=100$ kHz, $g_0/2\pi=1$ MHz, $\bar{\alpha}=0.02$ and $\beta =1.6$.}}
\end{figure*}
where $T_A$ is the partial transpose and $||.||_1$ is the trace norm. We obtain the dynamics of the reduced density matrix of the system by numerically solving Eq. \eqref{eq:eq4} with the following experimentally realizable parameters. The degenerate mechanical modes are characterized by their resonance frequency and decay rate $\omega_m/2\pi=15$ MHz and $\Gamma/2\pi=15$ Hz, respectively. The cavity is red-detuned with respect to the probe field by twice the mechanical frequency ($\Delta=-2\omega_m$), and its decay rate is $\gamma_a/2\pi=100$ kHz. The single photon optomechanical interaction strength $g_0/2\pi =1$ MHz. The steady state cavity field amplitude $\bar{\alpha} = 0.02$ and $\beta = \sqrt{|\varepsilon_d|/|g_2|}=1.6$.
\section{Results}
\label{sec:results}
    A general procedure for creating a mechanical cat state involves parametrically driving a bosonic mode in the presence of engineered strong nonlinear dissipation. From the equation of motion of the reduced density matrix for our model system, we obtain a parametric drive that does not affect a single bosonic mode, but instead drives the symmetric combination of bare mechanical modes $(b_1+b_2)$. Additionally, the nonlinear optomechanical interaction induces a pronounced quadratic dissipation channel for the mode $(b_1+b_2)$ surpassing that of the linear damping. Introducing bright and dark modes, the dynamical evolution of the reduced density matrix reduces to a phase-dependent parametrically driven two-phonon process. We consider the situation in which thermal phonons of each bare mechanical mode are absent. We numerically solve Eq. \eqref{eq:eq4} with a well-established QuTip solver. At $\theta = 0^\circ$, the evolution of the bright mode ($D_{+}$) creates a Schr\"{o}dinger cat state as shown in Fig. \ref{fig:fig2}(a).
\begin{figure*}[!ht]
	\centering
	\includegraphics[width=\textwidth]{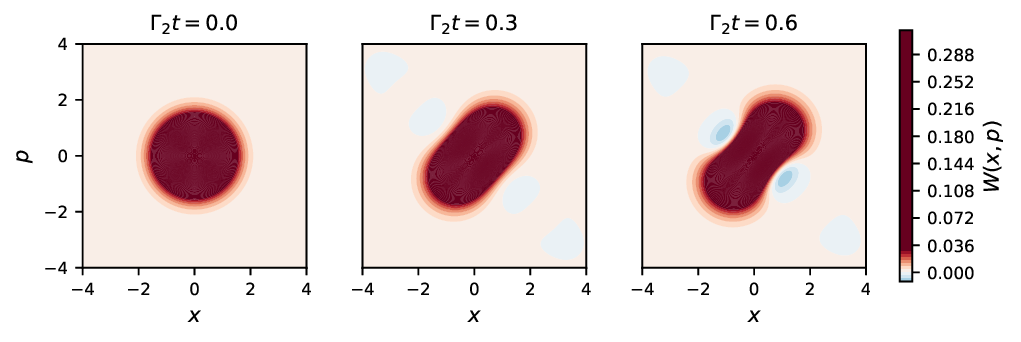}
	\caption{Temporal evolution of the conditional state of the bright mode. The phonon hopping interaction strength is characterized by $\chi/\Gamma_2=0.01$ and $\phi=\pi/2$. All the other parameters are same as in Fig.~\ref{fig:fig2}.}
    \label{fig:fig3}
\end{figure*}
However, for this particular value of $\theta$, the dark mode ($D_{-}$) becomes completely unsensitive to the parametric drive and is also decoupled from the quadratic dissipation channel. As a result, it remains Gaussian throughout the system's dynamics as presented in Fig. \ref{fig:fig2}(b). 
In contrast, when the interaction between the two mechanical modes is characterized by a phase $\theta=\pi$, the quadratic dissipation and a parametric drive act on $D_-$. It leads to a non-Gaussian cat state formation in the dark mode, while the bright mode remains completely decoupled from the nonlinear dynamics as seen from Fig. \ref{fig:fig2}(c)-(d). In other words, for $\theta = 0$ and $\pi$, the steady-state wave function $|\psi\rangle$ becomes proportional to $\left(|\alpha_1\rangle\pm|-\alpha_1\rangle\right)_{D_+}\otimes|0\rangle_{D_-}$ or $|0\rangle_{D_+}\otimes\left(|\alpha_1\rangle\pm|-\alpha_1\rangle\right)_{D_-}$, with $\alpha_1^2=-i\varepsilon_2/\Gamma_2$ respectively. Thus, the steady-state factorizes into a product of independent states in the $D_+$ and $D_-$ subspaces, indicating that there is no entanglement between them. This lack of entanglement can be further confirmed by examining the temporal evolution of the entanglement between the collective modes, as demonstrated in Fig. \ref{fig:fig2}.\\
The situation changes considerably when we examine the intermediate-phase value of the inter-mechanical-mode interaction. At $\theta = \pi/2$, the quadratic dissipation acts on bright and dark modes equally. We incorporate it into the evolution of the reduced density matrix for a weak Kerr interaction. In addition, we assume that the energy difference between the eigenmodes is much smaller than the two-phonon dissipation rate ($\chi/\Gamma_2<<1$) and neglect the lossy cavity-induced linear damping of the collective modes ($\Gamma_i/\Gamma_2<<1$), while accounting for the local damping of individual mechanical modes in the absence of thermal excitations. The resulting simplified master equation can be expressed as follows:
\begin{align}
\label{eq:eq8}
\dot{\rho_s}=&\Gamma_2\mathcal{L}\left[-i\left(D_++D_-\right)^2-\beta_0^2\right]\rho_s+\Gamma\left(\mathcal{L}[D_+]+\mathcal{L}[D_-]\right)\rho_s,
\end{align}
where $\beta_0^2=-2i\varepsilon_2/\Gamma_2$. The density matrix of the mechanical system at any transient time $t>t_0$ can be expressed in the collective mode basis as
\begin{align}
\label{eq:eq9}
\rho_{\textrm{a}}(t>t_0)&={\mathcal{N}(t)}
(|\alpha,\alpha\rangle\langle\alpha,\alpha|+|-\alpha,-\alpha\rangle\langle-\alpha,-\alpha|\nonumber\\
&+c(t)\left(|\alpha,\alpha\rangle\langle -\alpha,-\alpha|+|-\alpha,-\alpha\rangle\langle\alpha,\alpha|\right)),
\end{align}
where $c(t)=e^{-4|\alpha|^2}+\left(c(t_0)-e^{-4|\alpha|^2}\right)e^{-4\Gamma|\alpha^2|(t-t_0)}$ and $\mathcal{N}(t)=\left(2+2 ~c(t)~e^{-4|\alpha|^2}\right)^{-1}$ are the temporal evolution of the coherence and the normalization constant, respectively. The correctness of the density matrix is discussed in Appendix \ref{appendix2}. The first two terms of the analytical expression of the reduced density matrix describe the statistical mixture of the two coherent states \(|\alpha, \alpha\rangle\) and \(|-\alpha, -\alpha\rangle\) with $\alpha=(\beta_0/2)e^{i\pi/4}$. The last two terms contribute an off-diagonal coherence term, which leads to quantum interference that reaches its maximum at \(t = t_0\). This results in maximum bipartite entanglement at \(\Gamma_2 t = 0.5\) (represented by the black dashed line), as shown in Fig. \ref{fig:fig2}. Subsequently, it decoheres at a rate of \(\gamma_2 = 4\Gamma|\alpha|^2\). Furthermore, we assess the nonclassicality of the bipartite state in Eq. \eqref{eq:eq9} by a conditional measurement protocol. A projective measurement on the dark mode with an outcome $|\psi\rangle_{D_-}$ produces the conditional state of the bright mode as
\begin{align}
\label{eq:eq10}
\rho_{D_+}^{(\psi)}(t)=\langle\psi|\rho_a(t>t_0)|\psi\rangle_{D_-}.
\end{align}
For an analysis of the fock state with the outcome $n=0$, {\it{i.e.,}} $|\psi\rangle_{D_-}=|n=0\rangle_{D_-}$, the conditional state becomes
\begin{align}
\label{eq:eq11}
\rho_{D_+}^{(0)}(t)&= \mathcal{N}_1(t)(|\alpha\rangle\langle\alpha|+|-\alpha\rangle\langle-\alpha|\nonumber\\
&+c(t)(|\alpha\rangle\langle-\alpha|+|-\alpha\rangle\langle\alpha|)),
\end{align}
where $\mathcal{N}_1(t)=\mathcal{N}(t)e^{-|\alpha|^2}$ is the modified normalization constant. The presence of quantum interference is indicated by the Wigner negativity. We solve Eq. \eqref{eq:eq9} and calculate the Wigner function, considering the vacuum state as a measurement outcome on the dark collective mode as shown in Fig. \ref{fig:fig2}(a)-(c). The correctness of the approximated conditional state is evaluated from the Uhlmann fidelity
 \begin{align}
 \label{eq:eq12}
\mathcal{F}(t)=\left(\textrm{Tr}\sqrt{\sqrt{\rho_{D_+}^{(0)}(t)}\rho(t)\sqrt{\rho_{D_+}^{(0)}(t)}}\right)^2.
 \end{align}
Initially, the conditional state exhibits a Gaussian Wigner function with a fidelity of \(\mathcal{F} = 0.825\) when the target state is represented in Eq. \eqref{eq:eq11}. As time progresses, the probability distribution reveals negativity, which indicates an increase in fidelity, reaching values of \(\mathcal{F} = 0.97\) at \(\Gamma_2 t = 0.3\) and \(\mathcal{F} = 0.98\) at \(\Gamma_2 t = 0.6\).
\subsection{Bipartite entanglement}
\label{sec:bip_ent}
The previous section examines the evolution of the approximated reduced density matrix during the creation of a two-mode mechanical cat state in the absence of thermal excitations in each mechanical mode. 
\begin{figure}[h]%
	\centering
	{\includegraphics[width=0.47\textwidth]{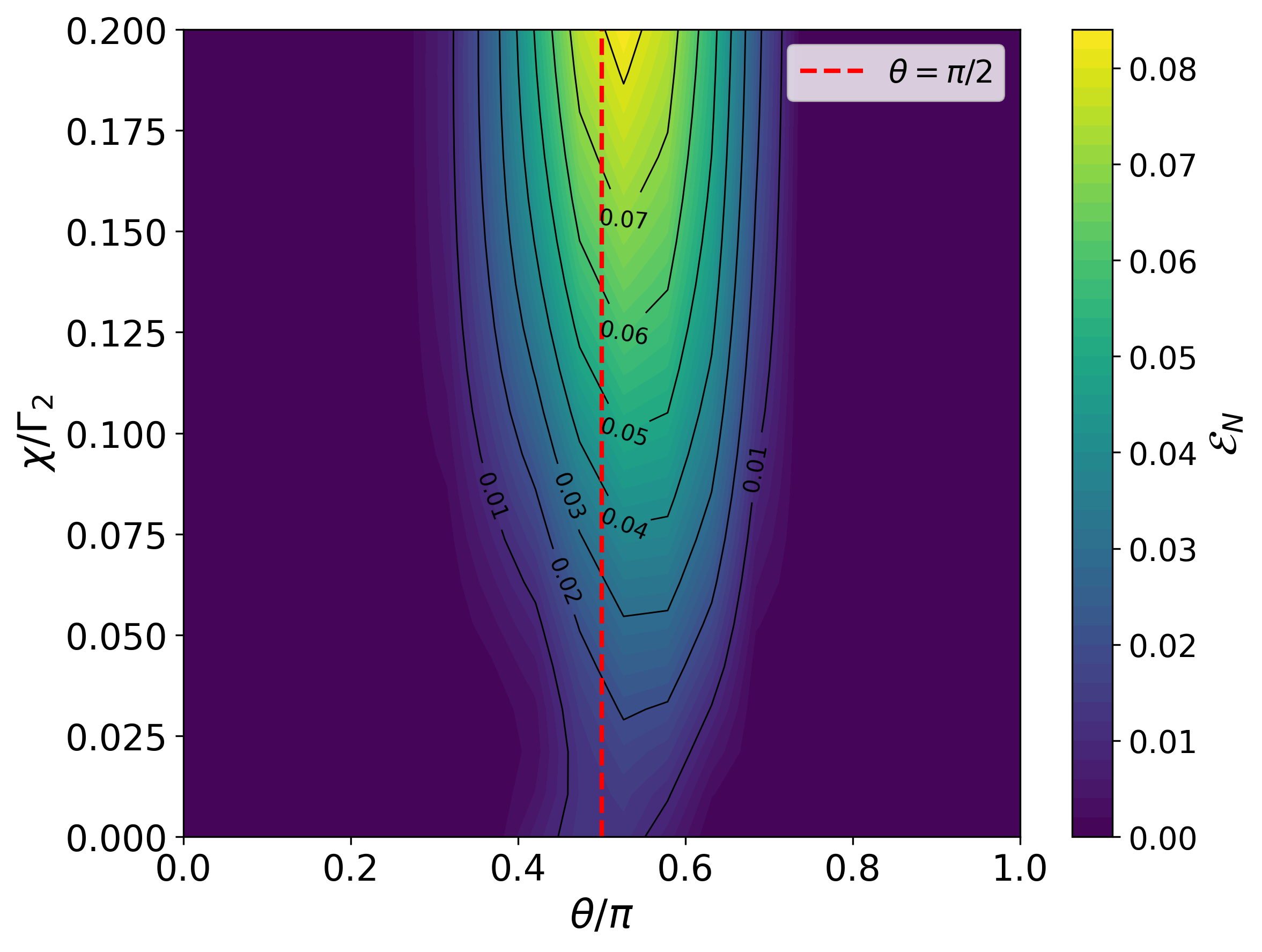}}
	\caption{\label{fig:fig3}{Steady-state bipartite entanglement between the collective modes as a function of $\chi$ and $\theta$, when thermal excitation of each mechanical mode is $0.5$. Red dashed line suggest $\theta=\pi/2$. All the other parameters are the same as Fig. \ref{fig:fig2}.}}
\end{figure}
This two-mode cat state is represented by the density matrix in Eq. \eqref{eq:eq9}, which describes a quantum superposition of the states \(|\alpha, \alpha\rangle\) and \(|-\alpha, -\alpha\rangle\). The presence of non-zero coherence, denoted as \(c(t)\), indicates that the reduced density matrix is not separable; that is, \(\rho_a(t) \neq \rho_{D_+} \otimes \rho_{D_-}\). Lack of separability signifies genuine bipartite entanglement between the constituent modes, particularly the collective modes in this context. The temporal evolution of \(\varepsilon_N\) shown in Fig. \ref{fig:fig2} indicates that coherence reaches its maximum at \(\Gamma_2 t_0 = 0.5\). 
Alongside the numerical calculation, we analytically evaluate the logarithmic negativity using Eq. \eqref{eq:eq9} and compare the results in tabular form. At the normalized time of \(0.5\), both methods show good agreement. However, at steady state, the two approaches yield significantly different results. This discrepancy suggests that the steady-state density matrix depends on the self-Kerr interaction, the dispersive coupling between the collective modes, and the phase-dependent phonon hopping interaction. Among these factors, the phonon hopping interaction strength $\chi$ and the phase $\theta$ are free parameters of the model.
\begin{table}[h]
\caption{\label{tab:table1} }
\begin{ruledtabular}
\begin{tabular}{ccd}
$\Gamma_2t$&From Eq. \eqref{eq:eq9}&
\multicolumn{1}{c}{From Eq. \eqref{eq:eq4}}\\
\hline
0.5&0.89&0.81\\
$\infty$&0.10&0.18\\
\end{tabular}
\end{ruledtabular}
\end{table}
Fig. \ref{fig:fig3} illustrates the effect of phonon hopping interaction on the robustness of steady-state entanglement in the presence of thermal phonon ($n_b = 0.5$). For weak phonon-hopping strength $\chi/\Gamma_2$, bipartite entanglement sustains only around $\theta=\pi/2$ (red-dashed line). In this regime, increasing $\chi$ leads to an enhancement of logarithmic negativity. Our numerical study is restricted to $n_b\leq0.5$ and $\chi/\Gamma_2\leq0.2$, where the numerical results remain converged with respect to Hilbert space truncation; that is, increasing the Hilbert-space dimension beyond a certain value does not produce any noticeable change in the results. However, for higher thermal phonons, determining the maximum Hilbert space truncation that yields converging numerical results may become challenging.
Since each of the collective modes depends on the bare mechanical modes, as presented in Eq. \eqref{eq:eq11}, we investigate bipartite entanglement between the bare mechanical modes, as shown in Fig. \ref{fig:fig4}. For any phase values between \(0\) and \(\pi\), these modes are weakly entangled in the presence of thermal excitation at \(0.5\). However, the amplitude and phase properties of the phonon-exchange interaction strength enable tunability of bipartite entanglement. The bare mechanical modes exhibit entanglement over a much wider range of \(\theta\) compared to the collective modes, reaching a maximum around \(\theta = \pi/2\). The results also predict that, in the regime of weak phonon interaction strength, the inseparability of bare mechanical modes increases with rising values of \(\chi\).
\begin{figure}[h]%
	\centering
	{\includegraphics[width=0.47\textwidth]{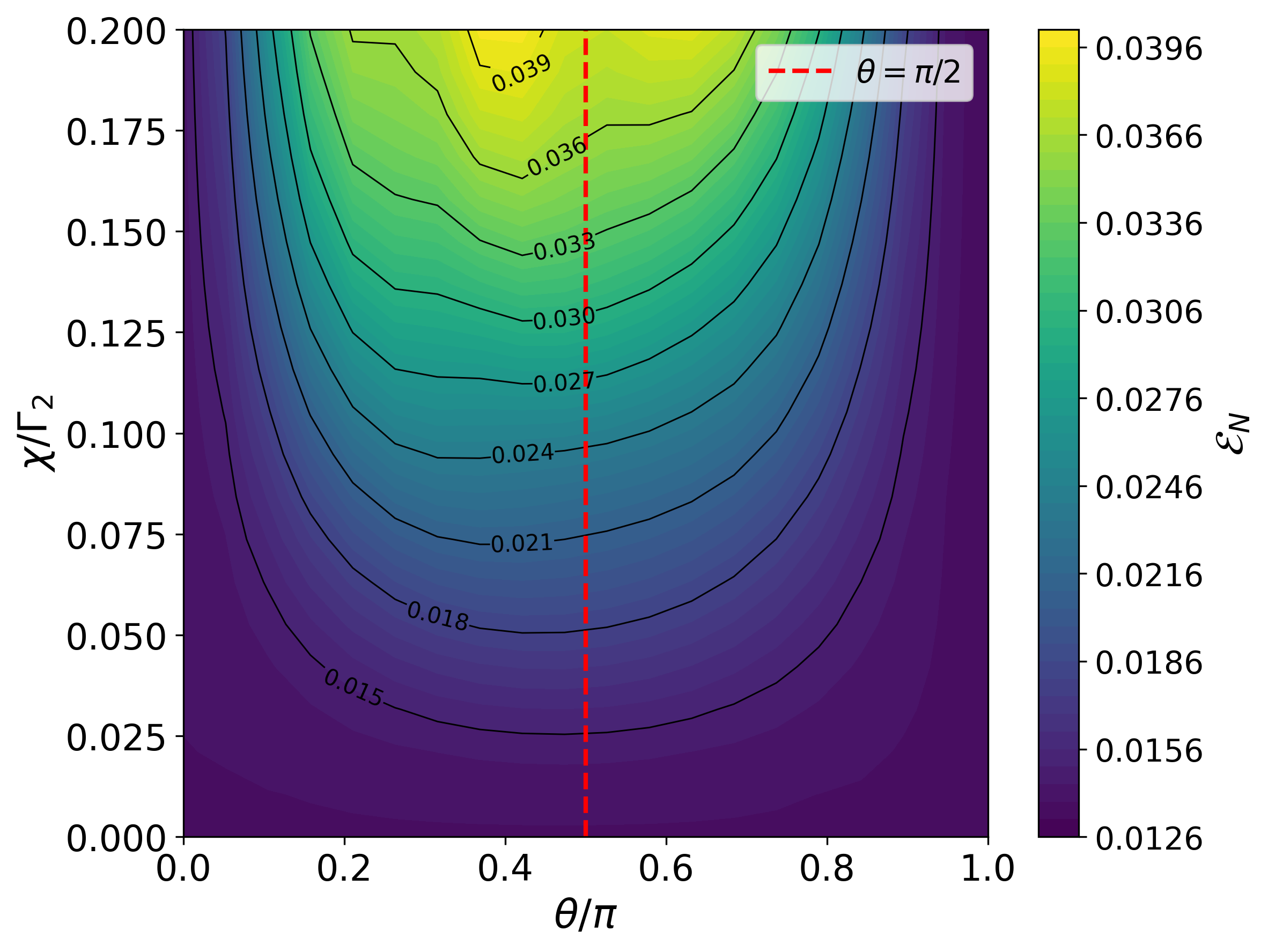}}
	\caption{\label{fig:fig4}{Steady-state bipartite entanglement between the bare mechanical modes as a function of $\chi$ and $\theta$, when thermal excitation of each mechanical mode is $0.5$. Red dashed line suggest $\theta=\pi/2$. All the other parameters are the same as Fig. \ref{fig:fig2}.}}
\end{figure}
\section{Conclusion}
\label{sec:conclusion}
In conclusion, we investigated the generation of a phase-dependent two mode Schr\"odinger cat state in a hybrid cavity optomechanical platform comprising a single cavity mode and two interacting degenerate mechanical modes. A bichromatic drive is applied to the lossy cavity. Under a specific cavity-detuning condition, adiabatic elimination of the cavity mode intensifies quadratic dissipation and induces parametric drive of $(b_1+b_2)$ mode in the presence of local damping. We show that when $\theta=0$, the SCS is formed in one of the collective modes, while for $\theta=\pi$, the SCS is generated in the other collective mode. A general value of $\theta$ produces a phase-dependent two-mode SCS. For a specific $\theta=\pi/2$, we determine the temporal evolution of the reduced density matrix. This indicates that the hopping phase directly influences the entanglement between the collective modes. Comparison of analytical predictions with numerical simulations highlights that neglected terms in the effective Hamiltonian significantly affect steady-state entanglement. Additionally, phase-dependent phonon-hopping interactions are shown to tune steady-state entanglement between bare mechanical modes in the presence of thermal noise. 

\section{Acknowledgements}
\label{sec:acknowledgements}
We acknowledge funding from the Okinawa Institute of Science and Technology. The authors acknowledge support from the Scientific Computing and Data Analysis (SCDA) section  at OIST and the use of the DEIGO supercomputing cluster.

\section{Data Availability}
\label{sec:data}
Data sharing is not applicable to this article as no datasets were generated or analyzed during the current study. Computer codes to reproduce the figures can be obtained from the corresponding author upon reasonable request.

\appendix
\begin{widetext}
\section{\label{appendix1}Reduced master equation for the mechanical modes}
To obtain the effective Hamiltonian up to the leading order in $g_0/\omega_{m_i}$, we utilize the Schrieffer-Wolff (SW) transformation $H' = e^s H_f e^{-s}$ with $S=\sum\limits_{i=1}^2(g_0/\omega_{m_i})a^\dagger a(b_i^\dagger-b_i)$. Under the SW transformation, the effective Hamiltonian can be written as
\begin{align}
\label{eq:eqA1}
\frac{H'}{\hbar}&=-\Delta a^\dagger a + \sum\limits_{i=1}^2\omega_{m}b_i^\dagger b_i+\sum\limits_{i=1}^2g_1\left(a+a^\dagger\right)(b_i+b_i^\dagger)
+\sum\limits_{i=1}^2g_2\left(a - a^\dagger\right)\left(b_i^2-{b_i^\dagger}^2\right)+\sum\limits_{i\not= j}\chi e^{i \theta_{ij}}b_i^\dagger b_j\nonumber\\
&+2g_2\left(a-a^\dagger\right)\left(b_1 b_2-b_1^\dagger b_2^\dagger\right)
+\left(\varepsilon_d a^\dagger e^{i\Delta_p t}+\textrm{h.c.}\right),
\end{align}
with second-order effective coupling strength $g_2=g_0^2\bar{\alpha}/\omega_m$. Finally, invoking the slowly moving envelope approximation as $a\sim a~e^{i\Delta t},~b_1\sim b_1~e^{-i\omega_mt},~b_2\sim b_2~e^{-i\omega_mt}$, we decompose the Hamiltonian as
\begin{align}
\label{eq:eqA2}
H'' =& H_0+H_1+H_2+H_3.
\end{align} 
 All the components of $H''$ are
\begin{widetext}
\begin{subequations}
\begin{align}
\label{eq:eqA3}
H_0/\hbar =& \chi\left(e^{i\theta}b_1^\dagger b_2+e^{-i\theta}b_1 b_2^\dagger\right),\\
H_1/\hbar=&\sum\limits_{i=1}^2\left(g_1^*ae^{i\Delta t}+g_1 a^\dagger e^{-i\Delta t}\right)\left(b_i e^{-i\omega_m t}+b_i^\dagger e^{i\omega_m t}\right),\\
H_2/\hbar=&\sum\limits_{i=1}^2\left(g_2^*ae^{i\Delta t}-g_2 a^\dagger e^{-i\Delta t}\right)\left(b_i^2 e^{-2i\omega_m t}-{b_i^\dagger}^2 e^{2i\omega_m t}\right)+2\left(g_2^* a e^{i\Delta t}-g_2a^\dagger e^{-i\Delta t}\right)\left(b_1b_2 e^{-2i\omega_mt}-b_1^\dagger b_2^\dagger e^{2i\omega_m t}\right),\\
H_3/\hbar=&\left(\varepsilon_d a^\dagger e^{i\Delta_c t}+\varepsilon_d^* a e^{-i\Delta_c t}\right).
\end{align} 
\end{subequations}
\end{widetext}
The detuning of the coherent drive with respect to the cavity resonance frequency is $\Delta_c=\omega_c-\omega_d$. We consider a scenario involving a bad cavity where the cavity decay rate $\gamma_a$ is much stronger than the enhanced optomechanical coupling strength ($g_1$), second-order effective coupling strength ($ g_2$), interaction strength between two degenerate mechanical modes ($\chi$), and mechanical damping rate ($\Gamma$). Under such a condition, we adiabatically eliminate the fast-decaying cavity mode following a well-established technique of open quantum systems \cite{Gonzalez-Ballestero2024Tutorial:Systems,azouit2016adiabaticeliminationopenquantum,10.1143/PTP.20.948,Wilson-Rae2008Cavity-assistedResonators}. The reduced master equation for the mechanical modes
\begin{align}
\label{eq:eqA4}
\dot{\rho}_s=\mathcal{L}_s\rho_s+\sum\limits_{i,j=1}^3 M_{ij},
\end{align}
where
\begin{align}
\label{eq:eqA5}
M_{ij}=&-\frac{1}{\hbar^2}\textrm{Tr}\left(\left[H_i(t),\int_0^\infty d\tau e^{\mathcal{L}_a\tau}\left[H_j(t-\tau),\rho_a\otimes\rho_{s}\right]\right]\right).
\end{align}
Substituting the coefficients $M_{ij}$ in Eq. \eqref{eq:eqA4}, the dynamics of the reduced density matrix of mechanical modes ($\rho_s=\rho_{b_1}\otimes\rho_{b_2}$), we obtain the dynamics of the reduced system as shown in Eq. \eqref{eq:eq4}.
\section{\label{appendix2}Correctness of the approximated density matrix}
In this section, we validate the analytically derived approximated density matrix outlined in Eq. \eqref{eq:eq9}, specifically in the regime where the phonon exchange interaction \(\chi\) is weaker than the two-phonon damping rate \(\Gamma_2\). We evaluate the Uhlmann fidelity
 \begin{align}
 \label{eq:appendix_1}
\mathcal{F}(t)=\left(\textrm{Tr}\sqrt{\sqrt{\rho_a(t>t_0)}\rho(t)\sqrt{\rho_a(t>t_0)}}\right)^2,
 \end{align}
where $\rho_a(t>t_0)$ and $\rho(t)$ are the target density matrix presented in Eq. \eqref{eq:eq9} and numerically evaluated density matrix at time $t$, respectively. Fig. \ref{fig:fig5} suggests the Uhlmann fidelity reaches a maximum around the normalized time $\Gamma_2t \approx 0.6$ for $\chi/\Gamma_2=0.01,~0.05,~\textrm{and}~0.10$. The solid black curve represents the fidelity remains constant at $0.97$ over a long temporal evolution when $\chi/\Gamma_2=0.01$. However, as \(\chi\) gradually increases, the fidelity experiences a decline. This observation confirms that the analytical density matrix presented in Eq. \eqref{eq:eq9} is accurate only under the condition of weak phonon exchange interaction. 
\begin{figure}[b!]%
	\centering
	{\includegraphics[width=0.47\textwidth]{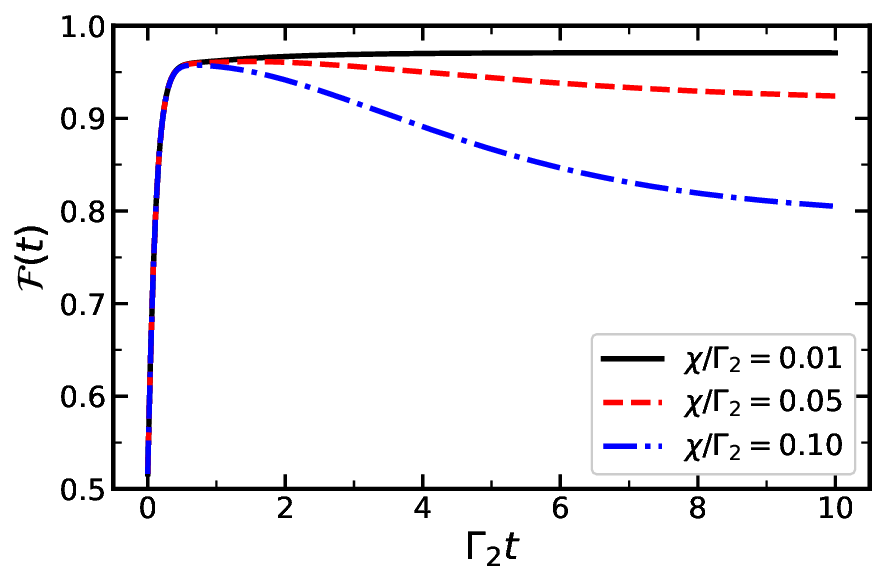}}
	\caption{\label{fig:fig5}{Temporal evolution of fidelity $\mathcal{F}(t)$ for $\chi/\gamma_2=0.01,~0.05,~0.1$. For all the the configurations, we consider $\theta=\pi/2$. All the other parameters are same as in Fig. \ref{fig:fig2}.}}
\end{figure}
\end{widetext}
\bibliography{apssamp, references}

\end{document}

%% file: model.pdf_tex
\begingroup%
  \makeatletter%
  \providecommand\color[2][]{%
    \errmessage{(Inkscape) Color is used for the text in Inkscape, but the package 'color.sty' is not loaded}%
    \renewcommand\color[2][]{}%
  }%
  \providecommand\transparent[1]{%
    \errmessage{(Inkscape) Transparency is used (non-zero) for the text in Inkscape, but the package 'transparent.sty' is not loaded}%
    \renewcommand\transparent[1]{}%
  }%
  \providecommand\rotatebox[2]{#2}%
  \newcommand*\fsize{\dimexpr\f@size pt\relax}%
  \newcommand*\lineheight[1]{\fontsize{\fsize}{#1\fsize}\selectfont}%
  \ifx\svgwidth\undefined%
    \setlength{\unitlength}{411.32319413bp}%
    \ifx\svgscale\undefined%
      \relax%
    \else%
      \setlength{\unitlength}{\unitlength * \real{\svgscale}}%
    \fi%
  \else%
    \setlength{\unitlength}{\svgwidth}%
  \fi%
  \global\let\svgwidth\undefined%
  \global\let\svgscale\undefined%
  \makeatother%
  \begin{picture}(1,0.8056914)%
    \lineheight{1}%
    \setlength\tabcolsep{0pt}%
    \put(0,0){\includegraphics[width=\unitlength,page=1]{model.pdf}}%
    \put(0.50015866,0.67452627){\color[rgb]{0.10196078,0.10196078,0.10196078}\makebox(0,0)[lt]{\lineheight{1.25}\smash{\begin{tabular}[t]{l}$a$\end{tabular}}}}%
    \put(0.10099332,0.10887025){\color[rgb]{0.10196078,0.10196078,0.10196078}\makebox(0,0)[lt]{\lineheight{1.25}\smash{\begin{tabular}[t]{l}$b_1$\end{tabular}}}}%
    \put(0.8628077,0.09369871){\color[rgb]{0.10196078,0.10196078,0.10196078}\makebox(0,0)[lt]{\lineheight{1.25}\smash{\begin{tabular}[t]{l}$b_2$\end{tabular}}}}%
    \put(0.19129716,0.40715354){\color[rgb]{0.10196078,0.10196078,0.10196078}\makebox(0,0)[lt]{\lineheight{1.25}\smash{\begin{tabular}[t]{l}$g_0$\end{tabular}}}}%
    \put(0.77558205,0.40830611){\color[rgb]{0.10196078,0.10196078,0.10196078}\makebox(0,0)[lt]{\lineheight{1.25}\smash{\begin{tabular}[t]{l}$g_0$\end{tabular}}}}%
    \put(0.45751737,0.13978097){\color[rgb]{0.10196078,0.10196078,0.10196078}\makebox(0,0)[lt]{\lineheight{1.25}\smash{\begin{tabular}[t]{l}$\chi,\theta$\end{tabular}}}}%
    \put(0.20357664,0.7634581){\color[rgb]{0.10196078,0.10196078,0.10196078}\makebox(0,0)[lt]{\lineheight{1.25}\smash{\begin{tabular}[t]{l}$\varepsilon_p,\omega_p$\end{tabular}}}}%
    \put(0.20473178,0.65416058){\color[rgb]{0.10196078,0.10196078,0.10196078}\makebox(0,0)[lt]{\lineheight{1.25}\smash{\begin{tabular}[t]{l}$\varepsilon_d,\omega_d$\end{tabular}}}}%
  \end{picture}%
\endgroup%